\documentclass[nonacm,acmsmall]{acmart}
\acmJournal{PACMPL}
\acmVolume{}
\acmNumber{} 
\acmArticle{}
\acmYear{2024}
\acmMonth{1}
\startPage{1}

\setcopyright{none}

\usepackage{listings}
\usepackage{placeins}
\usepackage{multicol}

\let\svthefootnote\thefootnote
\newcommand\freefootnote[1]{%
  \let\thefootnote\relax%
  \footnotetext{#1}%
  \let\thefootnote\svthefootnote%
}

\usepackage{xspace}

\newcommand{\todo}[1]{\textcolor{red}{{#1}}}

\newcommand{\toolname}{Leroy\xspace}

\newcommand{\code}[1]{\texttt{#1}}

\newcommand{\avgcompression}{1.04x\xspace}
\newcommand{\corpusSize}{122\xspace}

\newcommand{\ptwo}{$P_2$}

\usepackage{booktabs}   
\usepackage{subcaption}
\begin{document}

\title{Leroy: Library Learning for Imperative Programming Languages} 

\author{Abhiram Bellur}
\affiliation{
  \institution{University of Colorado, Boulder}            
  \country{USA}
}
\email{abhiram.bellur@colorado.edu}

\author{Razan Alghamdi}
\affiliation{
  \institution{University of Colorado, Boulder}            
  \country{USA} 
}
\email{razan.alghamdi@colorado.edu}

\author{Kidus Workneh}

\affiliation{
  \institution{University of Colorado, Boulder}            
  \country{USA}
}
\email{kidus.workneh@colorado.edu}      

\author{Joseph Izraelevitz}
\affiliation{
  \institution{University of Colorado, Boulder}
  \country{USA}
}
\email{joseph.izraelevitz@colorado.edu}

\begin{abstract}

Library learning is the process of building a library of common functionalities from a given set of programs. Typically, this process is applied in the context of aiding program synthesis: concise functions can help the synthesizer produce modularized code that is smaller in size. Previous work has focused on functional Lisp-like languages, as their regularity makes them more amenable to extracting repetitive structures.

Our work introduces \toolname, which extends existing library learning techniques to imperative higher-level programming languages, with the goal of facilitating reusability and ease of maintenance. \toolname wraps the existing Stitch framework for library learning and converts imperative programs
into a Lisp-like format using the AST.  Our solution uses
Stitch to do a top-down, corpus-guided extraction of repetitive expressions. Further, we prune abstractions that cannot be implemented in the programming language and convert the best abstractions back to the original language. 
We implement our technique in a tool for a subset of the Python programming language and evaluate it on a large corpus of programs. \toolname achieves a ompression ratio of \avgcompression of the original code base, with a slight expansion when the library is included. Additionally, we show that our technique prunes invalid abstractions.

\end{abstract}


\maketitle

\freefootnote{Presented at the 5th Intl.\ Wkshp.\  on Human Aspects of Types and Reasoning Assistants (HATRA).  Pasadena, CA, USA. 2024.}

\section{Introduction}
\label{intro}
Software engineers spend a large portion of their time improving the quality of their code. This includes refactoring, improving readability, fixing formatting issues, and adding documentation. In particular, engineers often refactor their code bases by abstracting reusable/repeated functionality from their code into library functions. These library functions (also called utility functions) capture parts of the program logic that are used by different software modules. 
Library extraction can be a tedious and error-prone process for developers if done manually. 

In prior work, Babble~\cite{Cao_2023babble} and Stitch~\cite{Bowers_2023stitch} leverage ideas from DreamCoder~\cite{ellis2020dreamcoder} to extract common functionality from code for library synthesis. However, these extraction tools are aimed at aiding the learning process of program synthesizers. When a program synthesizer learns from a compressed corpus, it is expected to learn to write smaller code that utilizes the abstracted libraries. This reduction of the synthesized output size reduces the possibility of errors. Prior work extracts functions over languages that use a functional Lisp-like syntax, with a focus on generating abstractions that provide the highest compression rate possible.

However, ideas from these techniques are unexplored for general-purpose programming languages, where developers would also like to extract common functionality for reusability and maintenance purposes. Particularly, reusability describes how well an extracted function can be reused throughout a given program corpus. Once a programmer has a good abstraction, they can simply reuse it in the future if they wish to elaborate their code. This extraction also affects maintenance in that a well-abstracted function can make the corpus more modularized, thus making it easier for programmers to maintain their code as fixing errors in the common functions propagate the changes of functionality through the entire corpus. 

In this work, we introduce \toolname, a technique that extends library extraction to general purpose programming languages by wrapping the the state-of-the-art Stitch~\cite{Bowers_2023stitch} library extraction tool. Our work targets a formal subset of Python, named \ptwo{}, as our chosen language.  \ptwo{} was designed as an exemplar language for teaching compiler classes~\cite{pythonbook}, and covers much of Python's functionality, including control, functions, and scoped variables. 

To leverage Stitch, \toolname first converts the original \ptwo{} program into its AST (abstract syntax tree), then
uses the AST to generate a Lisp-like program representation
acceptable to Stitch.  Stitch then generates candidate
functions for extraction; however, many of these candidates
may be illegal once converted back to Python.  Invalid abstractions could be macro-like, take invalid parameters, fail to return necessary variables, access variables out of scope, or simply be too small. We describe these issues in detail in Section~\ref{sec:design}. 

To evaluate \toolname, we test its compression abilities over a corpus of \corpusSize test programs from our university's compiler class. Additionally, we count the number of invalid abstractions removed by our technique and discuss the types of abstractions found by the tool.

Our paper proceeds as follows.  We first discuss the background on library synthesis in Section~\ref{sec:background}.  We then introduce our design of \toolname{} in Section~\ref{sec:design}, and evaluate in Section~\ref{sec:eval}.  We review related work in Section~\ref{sec:relwork}, and conclude with a discussion
of future work in Section~\ref{sec:conc}.


\section{Background}
\label{sec:background}

\subsection{Library Learning and Abstract Syntax Trees}
\begin{figure}[h]
  \centering
  \includegraphics[width=\linewidth]{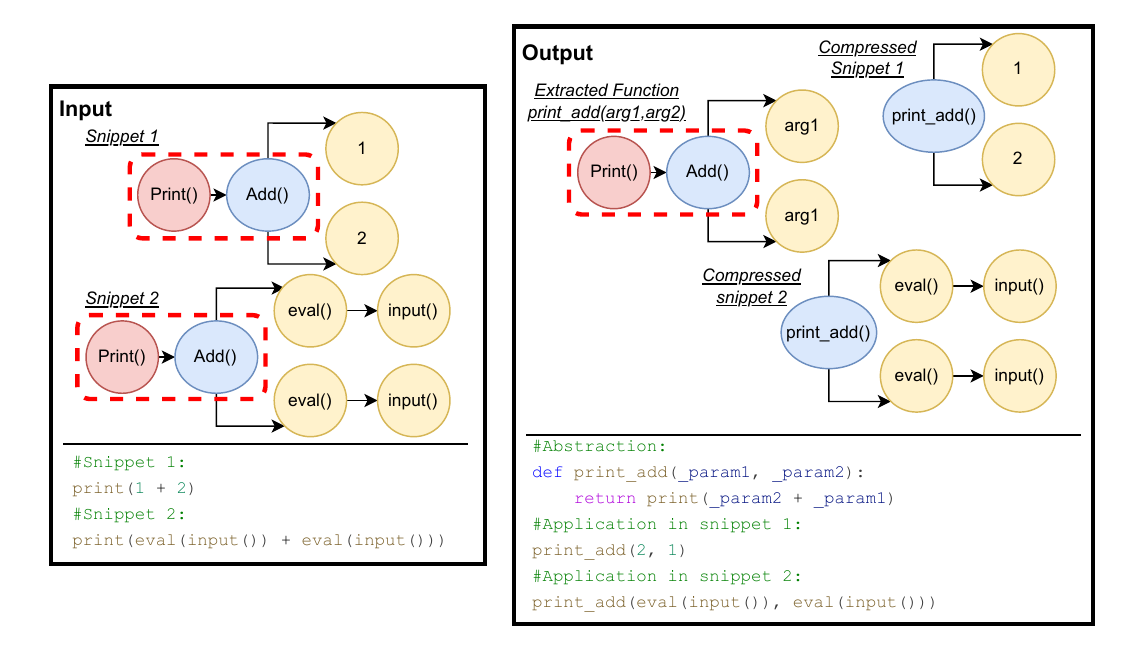}
  \caption{A library extraction example showing snippets of the input code, the output code, and their ASTs. The corpus (left) comprises two code snippets that share a common structure (highlighted). CTS can extract this common structure, here named \code{print\_add(arg1,arg2)}, and replace it with a function call (right).  }
  \label{figure:LibAbsAST}
\end{figure}

Library extraction involves identifying and extracting repeated functionalities from a corpus of programs, which is facilitated by analyzing the Abstract Syntax Trees (ASTs) of the programs. ASTs provide a structured representation of the code syntax, allowing algorithms to identify common patterns or partial subtrees that can be abstracted into reusable functions. Consider a corpus of programs where multiple programs contain a similar sequence of operations to print a sum of two things as shown on the left side of Figure~\ref{figure:LibAbsAST}. The library extraction process could identify this common pattern in the ASTs of these programs and abstract it into a reusable function \texttt{print\_add(n)}.

\subsection{Corpus Guided Top-down Search}
Corpus Guided Top-Down Search (CTS) is a recursive search method~\cite{Bowers_2023stitch} that looks for the optimal function, with respect to some utility measure, to be extracted from a given corpus. The algorithm starts with the search target of a partial program tree, naively an empty tree, and it searches the corpus for matching partial subtrees. The target's utility is then computed based on the number of matches found and the target's size, where larger targets with more matches have higher utility. In each subsequent round, the target is updated by adding child nodes to its leaf nodes. This expands the target but consequently reduces the matches. The children are added depending on the corpus as well as the expected utility. That is, the utility of expanding some branches of the target can be bounded by the current optimal utility, which guides the search to expand the target elsewhere. The search is concluded when the target reaches the optimal utility. The target then becomes the body of the extracted function, and the leaf's children that are not included in the extracted function become the function's arguments. Figure \ref{figure:LibAbsAST} shows an example of CTS function extraction.  


\section{Design}
\label{sec:design}
Stitch~\cite{Bowers_2023stitch} uses a corpus-guided top-down search approach directly, which works on the Lisp-like programs expected by Stitch, but the approach naively fails in less regular languages due to correctness issues: extracted common subtrees may not be valid in the original language.

For example, when Stitch is given a Python abstract syntax tree (AST) represented in a Lisp style, some elements or expressions, like the keywords "while" and "if", or operators like "==" and ">", are treated as first-class citizens of the programming language --- extracted subtrees could contain variables that hold such values, or functions could be called with these elements as parameters, resulting in syntactically incorrect calls, 
e.g. $compare(1,2,==)$ or assignments, e.g.\ \texttt{x = while}. Broadly, this issue arises because Stitch assumes a regular lambda calculus represented in a Lisp style, and a Python AST is not sufficiently regular. 
 
\toolname{} extends Stitch to correctly extract a library from a corpus of the language \ptwo{}, a large formal subset of Python 3 developed by Siek and Chang~\cite{pythonbook} in their textbook for compiler construction. All \ptwo{} programs are valid Python 3 programs with equivalent behavior.  The \ptwo{} language supports imperative-style programming, with scoped variable assignment, control, function calls, and nonnested definitions.  The language is dynamically typed on integers, booleans, lists, and dictionaries, with support for corresponding operations and comparisons.  Output is done through the \texttt{print} keyword, and input is handled via a fixed expression \texttt{eval(input())}, which takes user input, represented as an unsupported string type, and casts it to the appropriate type (e.g.\  if the user inputs \texttt{``True''}, the return value of the corresponding \texttt{eval(input())} expression will be the boolean \texttt{True} value).  The full grammar of the \ptwo{} we support is reproduced in the appendix --- we differ from the full published \ptwo{} only in dropping support for lambdas and nested function definitions, which remain as future work.

\begin{figure*}
  \includegraphics[width=\textwidth]{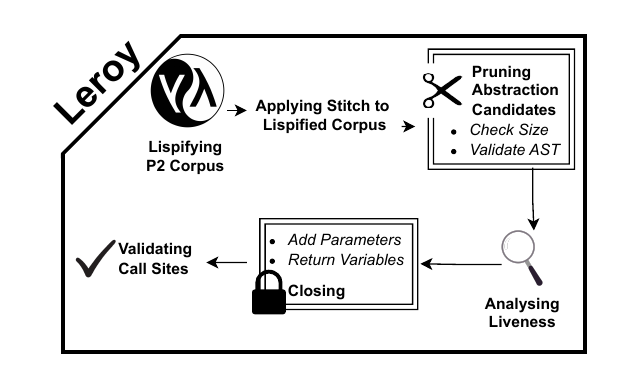}
  \caption{Architecture of the \toolname{} tool}
  \label{fig:design}
\end{figure*}

The architecture of our \toolname{} tool is shown in Figure ~\ref{fig:design}. As input,
\toolname{} expects in a corpus of \ptwo{} programs.  In the first step (\emph{lispify}), it converts the \ptwo{} programs into a Lisp-like format that is recognizable to Stitch~\cite{Bowers_2023stitch}.  Stitch then proposes a Lisp-like candidate for extraction.  The candidate is subsequently subjected to a series of \emph{pruning} checks that verify that the candidate can, indeed, be converted into a valid \ptwo{} function.  These checks correspond to a variety of issues that arise from Python's lack of regularization, and we describe them in depth in the following sections.  Once the candidate is verified, we use liveness analysis to add necessary parameters and return values --- this \emph{closing} is needed to handle the scope of live-in and live-out values from the candidate.  The final step is to \emph{validate call sites} to ensure that there is no naming collision between the function parameters and function contents, which could impact function behavior. 

In the following subsections, we cover each step of the \toolname{} pipeline in detail.

\subsection{Lispifying}
\label{sec:Lispify}
To use Stitch, \toolname must first convert the \ptwo{} program into a Lisp-like representation understandable to Stitch. 
This conversion treats AST nodes as nested function calls, where each node acts as a function and its child nodes as arguments. This approach enables \toolname to unfold the AST into a single, unified function call, simplifying the abstraction process.
For instance, consider a Python AST representing the addition operation \code{1+2}. 
The AST would look like a tree (see Figure~\ref{figure:LibAbsAST}, with an 'add' node at the root and the constants `1' and `2' as its children. 
In a functional style, we can represent the tree as a function call $add(1, 2)$, where the left and right children are the first and second operands respectively. 
In a Lisp-like form, this AST would appear as `(add 1 2)'. 

The other complication of lispifying is handling the imperative statements
of Python.  As an example, consider the following three statements \code{x=1; y=1; print(x+1)}. 
\toolname lispifies this sequence as $StatementList(x=1, StatementList(y=1, StatementList(print(x+y), \epsilon)))$, where $\epsilon$ is the empty statement. This encoding draws inspiration from the `let` statement in Lisp.

\subsection{Pruning Stitch's Abstractions}
Given the lispified code, Stitch suggests many candidate extracted functions that may not port well back to Python. To refine these candidates, \toolname prunes invalid candidates in a few categories. 

\subsubsection{Macro-Like Abstractions}
Stitch treats abstracted libraries effectively like macros, as it knows no better than to treat all program elements as first-class citizens. When a functionality is abstracted, it can simply be swapped with a call. This approach impacts program correctness as it results in incomplete abstractions, like abstracting ``\code{exp +}'' from ``\code{exp+exp}''. A potential abstraction ``\code{add(exp)}'' would replace an addition of ``\code{1+2}'' with the expression ``\code{add(1) 2}''. In Python, this is an invalid syntax. 

To identify and prune macro-like invalid candidates, \toolname attempts to reconstruct the candidate into a program AST form, then checks for the correctness of the yielded AST. ASTs are regenerated by inverting the lispify step from Section~\ref{sec:Lispify}, that is, the $StatementList$ nodes are converted back into lists, and special keyword nodes are replaced with their corresponding Python AST nodes (e.g.\ an $Add$ operator).  

Once the reconstructed AST is built, we verify the completeness of the AST by ensuring that all required child nodes of a parent node exist. Incomplete ASTs are indicative of macro-like structures, which are subsequently pruned. For example, an incomplete AST might involve a function call without required arguments, such as \code{add(a, )} where the second argument \code{b} is missing. 

\subsubsection{Invalid Parameters}
The search method of Stitch treats all tree nodes as potential ``holes'', that is, possible function parameters, which is incorrect in the case of Python whose AST nodes are not all first-class citizens. For example, when stitch abstracts over comparative sub-trees, like \code{x==y} or \code{x!=z} it can suggest abstractions that take three parameters, two operands, and a comparison operator. A potential candidate $f(x,op,y)$ would need to be called like: $f(x, ==, y)$ or $f(x, !=, y)$. As with macro-like abstractions, this candidate has invalid Python syntax.

To identify these cases, \toolname again analyzes the candidates’ body by converting it back into a Python AST and encoding parameters as identifier nodes. By validating these nodes against the AST structure, \toolname ensures the legitimacy of parameter usage within the candidate. For instance, in a comparison expression \code{(a > b)}, Stitch may produce a candidate of the form $compare(a, comparator, b)$ instead of $greater\_than(a,b)$.  Candidates that take invalid parameters are pruned out. 

\subsubsection{Presenting Non-trivial Abstractions}
Because Stitch optimizes over the abstracted function size relative to its use, it may abstract highly repeated single-line functionalities. Such functionality is found in abundance within the code-base and unfortunately results in functions that, for example, simply add two numbers.

To enhance the utility of suggested abstractions, \toolname filters out trivial cases by imposing a minimum size requirement. This criterion ensures that the extracted abstractions are sufficiently complex and meaningful to warrant inclusion. This enforces that trivial functions, like simple addition or printing, are not abstracted alone despite having a high utility due to their high usage frequency. 

\label{closing}
\subsection{Closing Functions}

\subsubsection{Determining the Return Value}

Because Stitch treats abstractions as macros, it does not consider variable scoping in abstractions, which impacts code correctness in a language like Python. For example, Stitch can abstract assignments without returning the assignment targets. That is, it can suggest functions that change the value of a variable but never return the changed variable. In Python, such assignment will create a new variable local to the abstracted (called) function instead of changing the value of the variable from the calling function. Although syntactically correct, this change impacts the program's behavior.

\toolname complements the abstractions generated by Stitch by ensuring a useful return value. To find what needs to be returned, \toolname performs a liveness analysis across each candidate call site, to determine what variables are live out of the abstraction body. 
It is worth noting that the abstraction's parameters may themselves be live-out and need to be returned. If no variables are live-out, \toolname reverts to returning the last variable or expression in the abstraction’s body. Such an approach helps with the functional correctness of the abstractions in the target programming language which may be used as sub-expressions at the call site. 

\subsubsection{Determining Additional Parameters} 
Stitch's abstractions may assume the presence of certain variables, even if they are not passed into the abstraction, due to the disconnect between Python's imperative style and the Lisp-like Stitch input language. 

It is important to pass these values as parameters to avoid uninitialized variables in the function body. To find these variables, \toolname uses the liveness analysis to determine variables that are not live into the abstraction's body. It is assumed that these variables exist in the calling scope, and they are added as parameters to the candidate.

\subsection{Validating Call Sites} 
A final complication arising from Python's imperative nature is a possible renaming issue that emerges between a given call site's arguments and the function body.

This problem is best illustrated with an example. Consider the candidate \code{print5(arg)} in Figure~\ref{fig:possible-invalid-function-call} on line~\ref{cd:candidate}. This function prints the value \texttt{arg} five times. Stitch identifies that this abstraction could used instead of the loop on line~\ref{cd:callsite} by simply using the function call \texttt{print5(x)} (line~\ref{cd:use}) --- but an application of the function here would be erroneous. 
This application is an invalid use because \texttt{x} is re-computed every iteration in the loop, but passing the value as \texttt{arg} computes it only once at the call site. 

The issue here is an implicit renaming of the symbol \texttt{x} when aliased into an argument, that is, the value of \texttt{x} is used as the renamed \texttt{arg} (at line~\ref{cd:print}), but the name \texttt{x} has been lost otherwise (at line~\ref{cd:for}) and is no longer connected to the associated value.  Again, this issue arises from Python scoping rules diverging from Lisp-style programs.
To detect such invalid call sites, \toolname analyzes each call site and its arguments, then validates that no names in the call site's arguments are referenced in the function, thereby preventing this name clash issue.

\begin{figure}
    \begin{lstlisting}
    def print5(arg):  # Verified Stitch candidate *@\label{cd:candidate}@*
        for x in range(5): *@\label{cd:for}@*
            print(arg) *@\label{cd:print}@*

    for x in range(5):   # Stitch identified usage (invalid) *@\label{cd:callsite}@*
        print(x)
        
    print5(x)   # Invalid candidate use (parameter and body overlap) *@\label{cd:use}@*
    \end{lstlisting}
    \caption{A valid function, which could be called erroneously}
    \label{fig:possible-invalid-function-call}
\end{figure}


\section{Evaluation}
\label{sec:eval}

We developed \toolname with Stitch commit number 3d352e6 
and we evaluate it on a 
MacBook Pro machine running a macOS Ventura operating system version 13.3. We utilized Python version 3.9.6.
To evaluate it, we used a corpus of \corpusSize small Python programs 
that adhere to the \ptwo{} grammar, each with an average length of 7 lines of code. The corpus is the complete suite of unit test programs used in our university's compiler course. This strategy gives the worst-case scenario for \toolname as the corpus is smaller than most production code with less repetitive content.

We attempted to squeeze the maximum number of abstractions out of Stitch. To do so, we allowed Stitch to run until it could no longer create any new abstractions. We discarded abstractions that call other abstractions, as these are typically small and trivial.

While applying \toolname with a minimum size threshold of 20 AST nodes, we found 6 valid abstractions.  These abstractions perform various assignment and printing operations, due to the nature of our corpus. For example, the abstracted function \texttt{assign\_print}, assigns the third and fourth parameters to the first and second parameters respectively. Subsequently, it prints the sum of the newly assigned first and second parameters. This abstraction was applied at three different sites. The found abstractions are listed in Figure~\ref{fig:leroy-abstractions}. In the search process, \toolname pruned 3123 macro-like abstractions, 32 abstractions with invalid parameters, and about 42000 abstractions which were smaller than 20 AST nodes. \toolname prunes so many small abstractions because Stitch assumes that each sub-tree of the AST is a candidate abstraction during its search phase.

We found that each of the found functions was applied to 2.5 call sites on average. We measured the compression ratio (original AST size: rewritten AST size), to determine the quality of the \toolname's learned library; indicating to us the factor by which the programs shrink. Overall, we achieved a compression ratio of \avgcompression, excluding the size of the extracted library. Including the library, the corpus grew by 1.2\% in terms of AST nodes. This expansion is because closing increases the size of the abstraction generated by Stitch, both in the abstraction's body and at its call site. In future work, we plan to explore larger real-world corpora and expect to uncover larger redundancies and achieve better compression.

\begin{figure*}   
    \begin{subfigure}[t]{0.45\textwidth}
        \begin{lstlisting}[language=Python, firstnumber=1]
def assign_print(_param0, _param1,
    _param2,_param3):
    _param1 = _param3
    _param0 = _param2
    return print(_param1 + _param0)

def swap_print(_param0, _param1,
    _param2, _param3):
    _param1 = _param3
    _param0 = _param2
    print(_param1)
    print(_param0)
    tmp = _param1
    _param1 = _param0
    _param0 = tmp
    print(_param1)
    return print(_param0)

def assign_print_eq(_param0, _param1):
    x = _param1
    y = _param0
    return print(x == y)
        \end{lstlisting}
        
        \label{fig:func3}
    \end{subfigure}\hfill
    \begin{subfigure}[t]{0.45\textwidth}
        \begin{lstlisting}[language=Python, firstnumber=1]
def multi_assign(_param1, _param2,
    _param3):
     x = 1
     y = 2
     z = _param3
     _param2 = _param1
     return (_param2, x, y, z)

def assign_add(_param0, _param1,
    _param2, _param3):
    if int(_param3):
        _param1 = _param1 + _param2
    else:
        _param1 = _param1 + _param0
    return _param1

def multi_assign_2(_param1, _param2,
    _param3, x):
    y = _param3 + x
    z = _param2 + y
    w = _param1 + z
    return (w, y, z)
    
        \end{lstlisting}
       
        \label{fig:func5}
    \end{subfigure}
   
    \vspace{-1em}
    \caption{Some of the abstractions found by \toolname}
    \label{fig:leroy-abstractions}
\end{figure*}


\section{Related Work}
\label{sec:relwork}

\subsection{Library Extraction}
Library extraction aims to compress code by extracting common functionalities of multiple functions into a single function that is used repeatedly. Program synthesizers use such libraries with the bigger goal of synthesizing programs (the generation of programs that solve a given problem). Most current state-of-the-art work on library extraction is based on DreamCoder~\cite{ellis2020dreamcoder}, a program synthesizer that adopts the wake-sleep algorithm~\cite{wake-sleep} to bootstrap inductive program synthesis from a small problem corpus represented with a Domain Specific Language (DSL). Specifically, DreamCoder uses an EC\textsuperscript{2} (Explore, Compress, Compile)~\cite{EC2} synthesis algorithm, where it abstracts libraries in one of its two sleep stages to explore potential extensions to the given corpus, then compresses the found abstractions in the next sleep stage to grow the library used for training the recognition neural network and synthesize programs in the wake stage of the next cycle. However, DreamCoder's success in solving problems in various domains, such as physics laws and text editing, comes with a performance cost. Subsequent work aimed to reduce this performance cost by introducing better library extraction mechanisms. 

For example, Babble~\cite{Cao_2023babble} proposes utilizing semantics-based abstractions, which produce better abstractions in a shorter time. However, Babble's efficiency comes with an additional overhead of providing a semantic equivalence list along with the input corpus. With this list, Babble can produce abstractions over syntactically different, yet semantically equivalent code.
 
Stitch~\cite{Bowers_2023stitch} on the other hand aims to efficiently abstract libraries by reducing the abstraction search space. It does so by defining a utility measure and performing a top-down search that optimizes utility using the branch-and-bound method, which eliminates search branches that are upper-bounded by a utility smaller than the current best utility. 

Stitch, Babble, and Dreamcoder are all designed to produce abstractions over functional (Lisp-like) programming languages, so \toolname extends and augments techniques from previous approaches to perform library extraction over imperative programming languages.

\subsection{Using Large Language Models}
Lilo~\cite{grand2024lilo} and ReGAL~\cite{stengeleskin2024regal} are works that utilize Large Language Models (LLMs) to aid in their program synthesis goals.

Lilo~\cite{grand2024lilo} is a program synthesizer that, like DreamCoder, works in a wake-sleep fashionl o utilizes LLM as a part of its synthesis phase and uses the Stitch compression algorithm to learn libraries. Further, Lilo uses LLMs once again to document information about the learned libraries, which helps in its subsequent synthesis phases. Lilo focuses on synthesis for functional programming languages, while \toolname extends to imperative programming languages like Python.

ReGAL~\cite{stengeleskin2024regal} is a program synthesizer that aims to generate Python programs for a specific domain (such as drawing shapes) given a prompt. It uses LLMs in two phases: library extraction and program synthesis. In its first phase, it takes in a batch of prompts and programs in the task domain and utilizes an LLM to learn helper functions (library extraction). Then, given a new prompt, ReGAL asks an LLM to synthesize a program by providing the previously generated helper functions. While ReGAL can generate abstractions over a language like Python, its reliance on LLMs means its abstractions may not be provably equivalent to the original code. Further, the use of LLMs does not help it scale to the level of large software code bases, as LLMs are bounded by the size of their context window.  This is not a hindrance for \toolname, which is capable of processing a large corpus of programs with equivalent abstractions.

Additionally, while Lilo and ReGAL focus on program synthesis (with abstraction as a guiding step), \toolname focuses solely on learning abstractions over the corpus.

\subsection{Visual Abstraction, Program Learning, and Rewriting}
Library extraction tools, including \toolname{}, can be placed in a larger research area broadly focused on program understanding and rewriting.
For instance, other works cover library extraction with a focus on the visual representation of programs~\cite{jones2023shapecoder, wang2021learningVisAbst, Jones_2021}. ShapeCoder~\cite{jones2023shapecoder}  for example uses neural networks and e-graphs to not only extract useful abstractions from the visual modeling of programs but also explain input shapes using the abstractions. 
Other lines of work, despite not sharing the library extraction focus, share components of programming and natural language processing and recognition. This includes work on program learning~\cite{cropper2019playgol, DBLP:journals/corr/abs-2004-09931refproginduc,wong2022leveraging, demo, iyer2019learning,hocquette2024learning} and program rewriting~\cite{brandfonbrener2024verified, DBLP:conf/sat/NotzliRBNPBT19rewrite, ganeshan2023improving}, which use relevant methods such as finding idioms, and search algorithms such as the Monte Carlo tree search.

\section{Discussion and Future Work}
\label{sec:conc}

Although \toolname generalizes Stitch \cite{Bowers_2023stitch} and addresses 
the main issues Stitch has when fed with Python code, it has yet to be more extensively and rigorously tested. \toolname can also be extended further to better address 
the stated problems (lambdas, nested functions, classes) with more advanced program analysis methods. Furthermore, \toolname extends Stitch without utilizing program semantics as Babble~\cite{Cao_2023babble} does. Hence, \toolname can be enhanced by adopting from Babble and utilizing semantic equivalence to expand the abstraction choices. Lastly, currently \toolname is only tested for compression, but library extraction for Python also targets readability and other program quality metrics that can be encoded in the utility function, and are yet to be tested.

\bibliography{references}

\appendix
\FloatBarrier
\newpage
\section{Appendix: \ptwo{} Grammar} 
\label{sec:grammar}
\FloatBarrier
\toolname operates on a subset of Python 3 named \ptwo{} as described in ``A Problem Course in Compilation: From Python to x86 Assembly''~\cite{pythonbook}. \ptwo{} consists of integer and boolean values and variables, addition and subtraction, boolean operations, comparators, conditionals, ternaries, while loops, and function calls and definitions. Although the published \ptwo{} also supports nested function definitions and lambdas, we leave the support for these features to future work.  The grammar of our supported \ptwo{} is given below.

\begin{figure}[h]
\begin{lstlisting}
program ::= module
module ::= statement+
statement ::= simple_statement | compound_stmt
simple_statement ::= "print" "(" expression ")"
                   | target "=" expression
                   | expression
                   | "return" expression
                   | name "=" expression
expression ::= name | decimalinteger
             | "True" | "False"
             | "-" expression
             | "not" expression
             | expression "+" expression
             | expression "and" expression
             | expression "or" expression
             | expression "=="
             | expression "!="
             | expression "if" expression "else" expression
             | expression "is" expression
             | "(" expression ")"
             | "[" expr_list "]"
             | "{" key_datum_list "}"
             | subscription
             | expression "[" expression "]"
             | expression "(" expr_list ")"
             | "eval" "(" "input" "(" ")" ")"
subscription ::= expression "[" expression "]"
target ::= identifier | subscription
key_datum ::= expression ":" expression
key_datum_list ::= key_datum | key_datum "," key_datum_list
expr_list ::= expression | expression "," expr_list
id_list ::=identifier | identifier "," id_list
compound_stmt ::= "def" identifier "(" id_list ")" ":" suite
suite ::= "\n" INDENT statement+ DEDENT
name ::= identifier
decimalinteger ::= [0-9]+
\end{lstlisting}
\label{fig:grammar}
\caption{\ptwo{} Grammar}
\end{figure}

\end{document}